\documentclass[slac_one]{revtex4}
\newcommand{\RS}{{\rm{RS}}}
\newcommand{\R}{{\cal{R}}}
\usepackage{graphicx}
\usepackage{fancyhdr}
\begin{document}

\title{Effective Charges, Event Shapes and Power Corrections}

%

\author{C.J. Maxwell}
\affiliation{Institute for Particle Physics Phenomenology (IPPP), Durham University, U.K.}

\begin{abstract}
We introduce and motivate the method of effective charges, and consider
how to implement an all-orders resummation of large kinematical logarithms
in this formalism. Fits for QCD $\Lambda$ and power corrections are performed
for the ${e}^{+}{e}^{-}$ event shape obesrvables 1-thrust and heavy-jet mass,
and somewhat smaller power corrections found than in the usual approach 
employing the ``physical scale'' choice.
\end{abstract}

\maketitle

\thispagestyle{fancy}


\section{Introduction}
In this talk I will describe some recent work together with
Michael Dinsdale concerning the relative size of non-perturbative
power corrections for QCD event shape observables \cite{r1}.   
For ${e}^{+}{e}^{-}$ event shape {\it means} the DELPHI collaboration
have found in a recent analysis that, if the next-to-leading order (NLO)
perturbative corrections are evaluated using the method of effective
charges \cite{r2}, then one can obtain excellent fits to data without including any power corrections \cite{r3}. 
In contrast fits based on the use of standard fixed-order perturbation theory in the $\overline{MS}$ scheme with a physical choice
of renormalization scale equal to the c.m. energy, require additional power corrections
${C}_{1}/Q$ with ${C}_{1}\sim 1\;\rm{GeV}$. Power corrections of this size
are also predicted in a model based on an infrared finite coupling \cite{r4}
, which is able to fit the data reasonably well in terms of a single
parameter. Given the DELPHI result it is interesting to consider how
to extend the method of effective charges to event shape {\it distributions}
rather than means.\\

\section{The method of effective charges}
Consider an ${e}^{+}{e}^{-}$ observable ${\cal{R}}(Q)$, e.g. an event
shape observable- thrust or heavy-jet mass, $Q$ being the c.m. energy.
\begin{equation}
{\cal{R}}(Q)=
a(\mu,\RS)+r_1(\mu/Q,\RS) a^2(\mu,\RS) +
+ r_2(\mu/Q,RS)a^3(\mu,\RS) + \cdots,
\end{equation}
Here $a\equiv{\alpha}_{s}/\pi$. Normalised with the leading coefficient
unity, such an observable is called an {\it effective charge}. The couplant $a(\mu,\RS)$ satisfies
the beta-function equation
\begin{equation} 
\frac{da(\mu,\RS)}{d\ln(\mu)}=\beta(a)=-b a^2 (1 + c a + c_2 a^2 + c_3 a^3 + \cdots)\;.
\end{equation}
Here $b=(33-2{N}_{f})/6$ and $c=(153-19{N}_{f})/12b$ are universal,
the higher coefficients ${c}_{i}$, $i\ge{2}$, are RS-dependent and may be used to label the
scheme, together with dimensional transmutation parameter $\Lambda$ \cite{r5}.
The {\it effective charge} ${\cal{R}}$ satisfies the equation        
\begin{equation}
\frac{d\R(Q)}{d\ln(Q)}=\rho(\R(Q))=-b \R^2 (1 + c \R + \rho_2 \R^2 + \rho_3 \R^3 + \cdots)\;.
\end{equation}
This corresponds to the beta-function equation in an RS where the
higher-order corrections vanish and ${\cal{R}}=a$, the beta-function coefficients in this scheme
are the RS-invariant combinations  
\begin{eqnarray}
\rho_2 & = & c_2 + r_2 - r_1 c - r_1^2 
\nonumber \\
\rho_3 & = & c_3 + 2r_3 - 4r_1 r_2 - 2 r_1 \rho_2 - r_1^2 c + 2r_1^3.\;.
\end{eqnarray}
Eq.(3) for $d{\cal{R}}/d{\ln{Q}}$ can be integrated to give
\begin{equation}
b\ln\frac{Q}{{\Lambda}_{\cal{R}}}=
\frac{1}{{\cal{R}}}+c{\ln}\left[\frac{c{\cal{R}}}{1+c{\cal{R}}}\right]+
\int_{0}^{{\cal{R}}(Q)}{dx}\left[\frac{b}{\rho(x)}+\frac{1}{{x}^{2}(1+cx)}\right]\;.
\end{equation}
The dimensionful constant ${\Lambda}_{\cal{R}}$ arises as a constant of integration. It is related to the dimensional transmutation parameter
 ${\tilde{\Lambda}}_{\overline{MS}}$ by the exact relation,  
\begin{equation}
{\Lambda}_{\cal{R}}={e}^{r/b}{\tilde{\Lambda}}_{\overline{MS}}\;.
\end{equation}
Here
${r}\equiv{r}_{1}(1,\overline{MS})$ with $\mu=Q$, is the NLO perturbative
coefficient. Eq.(5) can be recast in the form
\begin{equation}
{\Lambda}_{\overline{MS}}=Q{\cal{F}}({\cal{R}}(Q)){\cal{G}}({\cal{R}}(Q)){e}^{-r/b}{(2c/b)}^{c/b}\;.
\end{equation}
The final factor converts to the standard convention for $\Lambda$. Here ${\cal{F}}({\cal{R}})$
is the {\it universal} function
\begin{equation}
{\cal{F}}({\cal{R}})={e}^{-1/b{\cal{R}}}{(1+1/c{\cal{R}})}^{c/b}\;,
\end{equation}
and ${\cal{G}}({\cal{R}})$ is 
\begin{equation}
{\cal{G}}({\cal{R}})=1-\frac{{\rho}_{2}}{b}{\cal{R}}+O({\cal{R}}^{2})+{\ldots}\;.
\end{equation}
Here ${\rho}_{2}$ is the NNLO ECH RS-invariant. If only a NLO calculation is available,
as is the case for ${e}^{+}{e}^{-}$ jet observables, then ${\cal{G}}({\cal{R}})=1$, and
\begin{equation}
{\Lambda}_{\overline{MS}}=Q{\cal{F}}({\cal{R}}(Q)){e}^{-r/b}{(2c/b)}^{c/b}\;.
\end{equation}
Eq.(10) can be used to convert the measured data for the observable
${\cal{R}}$ into a value of ${\Lambda}_{\overline{MS}}$ bin-by-bin. Such an
analysis was carried out in Ref. \cite{r6} for a number of ${e}^{+}{e}^{-}$
event shape observables, including thrust and heavy jet mass which we shall
focus on here. It was found that the fitted $\Lambda$ values exhibited
a clear plateau region, away from the two-jet region, and the region approaching
$T=2/3$ where the NLO thrust distribution vanishes. The result for 1-thrust
corrected for hadronization effects is shown in Fig. 1.
\begin{figure}
\begin{center}
\includegraphics[scale=1.0]{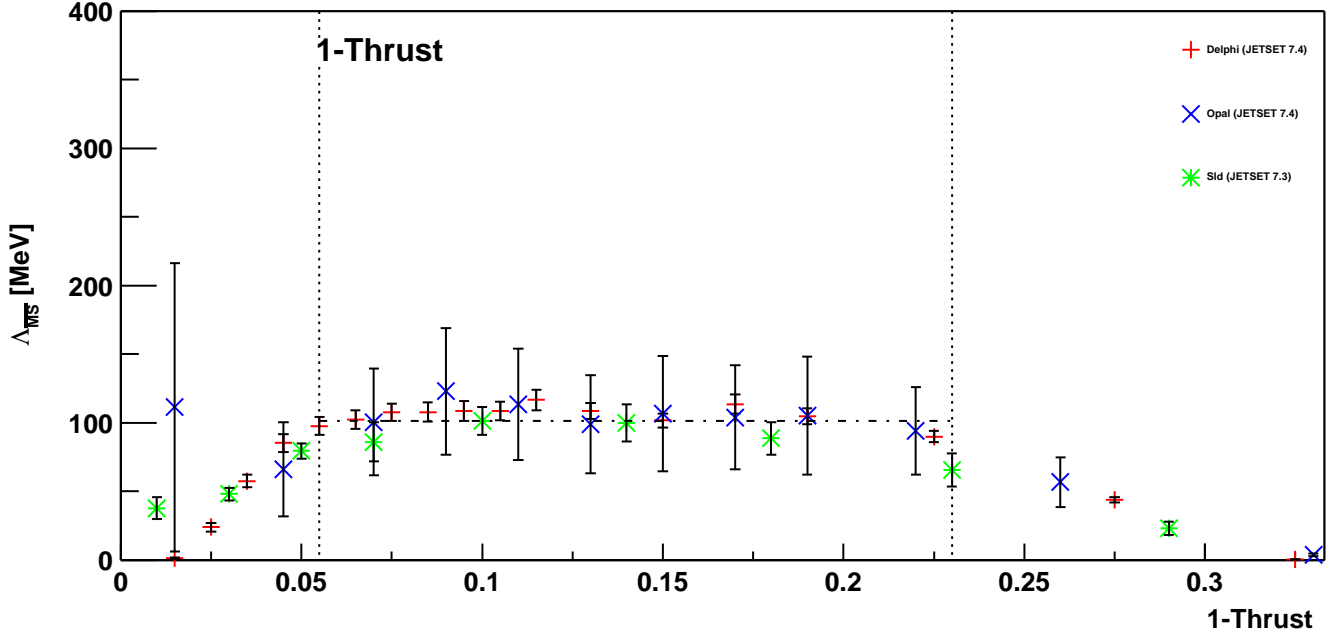}
\end{center}
\caption{Values of ${\Lambda}_{\overline{MS}}$ obtained for hadronization corrected 1-thrust data. \cite{r6}}
\end{figure}

Another way of motivating the effective charge approach is the idea of
``complete renormalization group improvement'' (CORGI) \cite{r6a}.
 One can write the NLO coefficient ${r}_{1}(\mu)$ as
\begin{equation}
{r_1}({\mu})=b{\ln}\frac{\mu}{{\tilde{\Lambda}}_{\overline{MS}}}-b{\ln}\frac{Q}{{\Lambda}_{\cal{R}}}\;.
\end{equation}
Hence one can identify scale-dependent $\mu$-logs and RS-invariant ``physical'' UV $Q$-logs.
Higher coefficients are polynomials in ${r}_{1}$.
\begin{eqnarray}
{r_2}&=&{r}_{1}^{2}+{r}_{1}c+({\rho}_{2}-{c_2})
\nonumber \\
{r_3}&=&{r}_{1}^{3}+\frac{5}{2}c{r}_{1}^{2}+(3{\rho_2}-2{c_2}){r_1}+(\frac{{\rho}_{3}}{2}-\frac
{c_3}{2})\;.
\end{eqnarray}
Given a NLO calculation of ${r}_{1}$, parts of ${r}_{2},{r_3},\ldots$ are ``RG-predictable''.
One usually chooses ${\mu}=xQ$ then $r_1$ is $Q$-independent, and so are all the $r_n$. The $Q$-dependence of 
${\cal{R}}(Q)$ then comes entirely from the RS-dependent coupling $a(Q)$. 
However, if we insist
that $\mu$ is held constant {\it independent of $Q$} the only $Q$-dependence resides in the ``physical''
UV $Q$-logs in $r_1$. Asymptotic freedom then arises only if we resum these $Q$-logs
to {\it all-orders}.
Given only a NLO calculation, and assuming for simplicity that that we have a trivial
one loop beta-function ${\beta}(a)=-b{a}^{2}$ so that $a(\mu)=1/b{\ln}(\mu/{\tilde{\Lambda}}_{\overline{MS}})$ the RG-predictable terms will be
\begin{equation}
{\cal{R}}=a({\mu})\left(1+{\sum_{n>0}}{(a({\mu}){r}_{1}({\mu}))}^{n}\right)\;.
\end{equation}
Summing the geometric progression one obtains
\begin{eqnarray}
{\cal{R}}(Q)&=&a({\mu})/\left[1-\left(b{\ln}\frac{{\mu}}{{\tilde{\Lambda}}_{\overline{MS}}}
-b{\ln}\frac{Q}{{\Lambda}_{\cal{R}}}\right)a({\mu})\right]
\nonumber \\
&=&1/b{\ln}(Q/{\Lambda}_{\cal{R}})\;,
\end{eqnarray}
The $\mu$-logs ``eat themselves'' and one arrives at the NLO ECH result
${\cal{R}}(Q)=1/b{\ln}(Q/{\Lambda}_{\cal{R}})$.\\

As we noted earlier, and as will be discussed by Klaus Hamacher in his talk
\cite{r6b}, use of NLO effective charge perturbation theory (Renormalization
Group invariant (RGI) perturbation theory) leads to excellent fits for
${e}^{+}{e}^{-}$ event shape {\it means} consistent with zero power
corrections, as illustrated in Figure 2. taken from Ref.\cite{r3}.
\begin{figure}
\begin{center}
\includegraphics[scale=0.6]{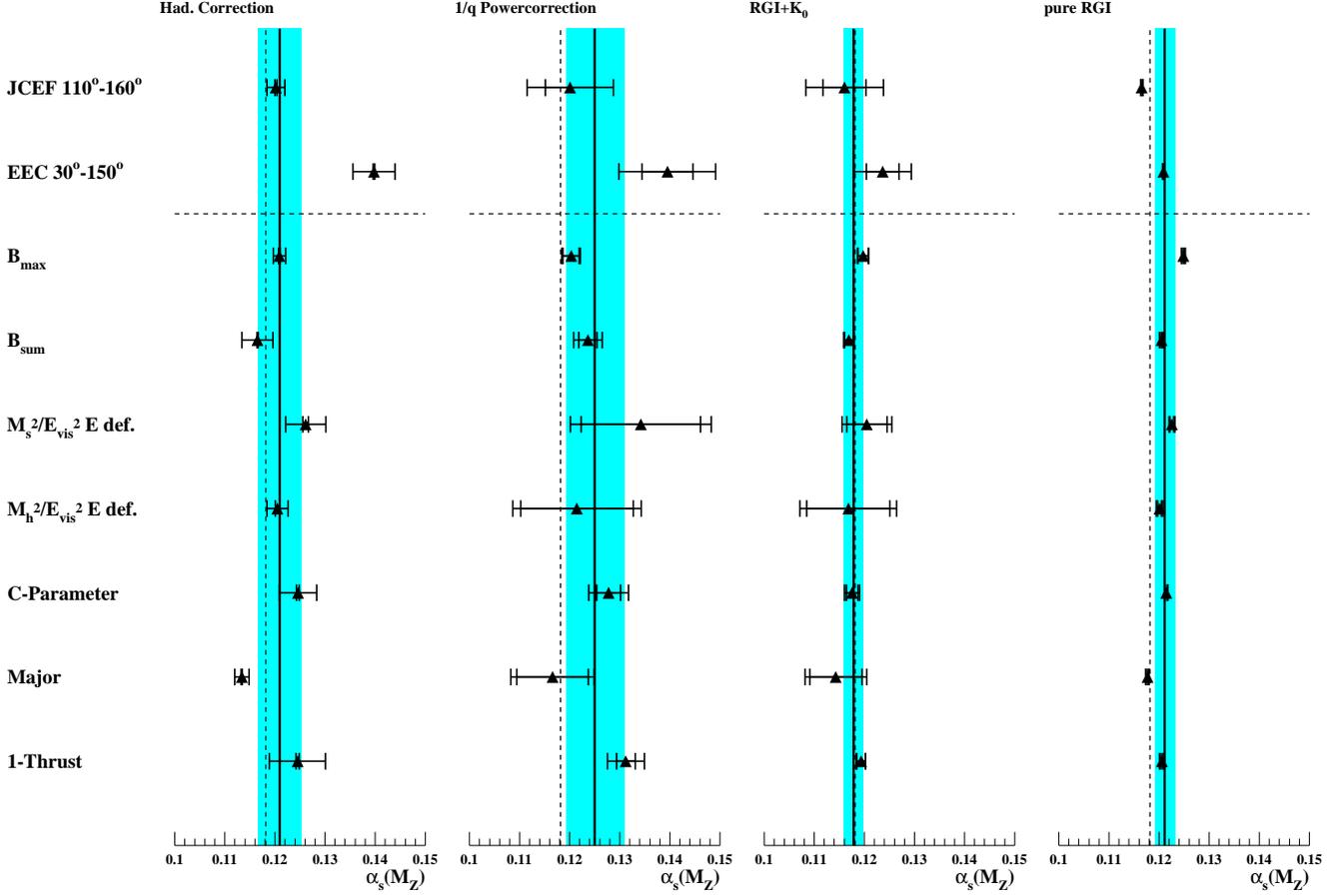}
\end{center}
\caption{Fits for ${\alpha}_{s}(M_Z)$ for means of ${e}^{+}{e}^{-}$ event shape
observables taken from Ref.\cite{r3}. The quality of the ``pure RGI'' fits
on the right is noteworthy.} 
\end{figure}
Given this result it would seem worthwhile to
extend the effective charge approach to event shape {\it distributions}. 
It is commonly stated that the method of effective charges is
inapplicable to exclusive quantities which depend on multiple scales. However given an observable
${\cal{R}}({Q}_{1},{Q}_{2},{Q}_{3},\ldots,{Q}_{n})$ depending on $n$ scales it can always be
written as 
\begin{equation}
{\cal{R}}={\cal{R}}({Q}_{1},{Q}_{2}/{Q}_{1},\ldots,{Q}_{n}/{Q}_{1}){\equiv}{\cal{R}}_{{x}_{2}{x}_{3}\ldots{x}_{n}}({Q}_{1})\;.
\end{equation}
Here the ${x}_{i}{\equiv}{Q}_{i}/{Q}_{1}$ are {\it dimensionless} quantities that can be
held fixed, allowing the ${Q}_{1}$ evolution of ${\cal{R}}$ to be obtained as before. In the 2-jet region
for ${e}^{+}{e}^{-}$ observables large logarithms $L={\ln}(1/{x}_{i})$ arise and need to be resummed to all-orders.

\section{ Resumming large logarithms for event shape distributions}
Event shape distributions for thrust ($T$) or heavy-jet mass
(${\rho}_{h}$) contain large kinematical logarithms, $L={\ln}(1/y)$, where 
$y=(1-T),\;{\rho}_{h},\cdots$. 
\begin{equation}
\frac{1}{\sigma} \frac{d\sigma}{dy} = A_{LL}(aL^2) + L^{-1} A_{NLL}(aL^2) + \cdots\;.
\end{equation}
Here $LL$, $NLL$, denote leading logarithms, next-to-leading logarithms, etc. For thrust
and heavy-jet mass the distributions {\it exponentiate} \cite{r7} 
\begin{eqnarray}
R_y(y')& \equiv& \int_0^{y'} dy \frac{1}{\sigma} \frac{d\sigma}{dy}
= C(a\pi) \exp(L g_1(a\pi L) 
\nonumber \\
&+& g_2(a\pi L) 
+ a g_3(a\pi L) + 
+ \cdots) + D(a\pi ,y)\;.
\end{eqnarray}
Here $g_1$ contains the LL and $g_2$ the NLL. $C=1+O(a)$ is independent of $y$, and
$D$ contains terms that vanish as $y\rightarrow{0}$.  
It is natural to define an effective charge ${\cal{R}}
(y')$ so that
\begin{equation}
R_y(y') = \exp(r_0(y'){\cal{R}}(y'))\;.
\end{equation}
This effective charge will have the expansion
\begin{equation}
r_0(L){\cal{R}}(L) = r_0(L) (a + r_1(L) a^2 + r_2(L) a^3 + \cdots)\;.
\end{equation}
Here ${r}_{0}(L)\sim{L}^{2}$, and the higher coefficients ${r}_{n}(L)$ have the
structure
\begin{equation}
r_n = r_n^{\rm LL} L^n + r_n^{\rm NLL} L^{n-1} + \cdots
\end{equation}
Usually one resums these logarithms to all-orders using the known closed-form expressions
for ${g}_{1}(aL)$ and ${g}_{2}(aL)$, where $a$ is taken to be the ${\overline{MS}}$ coupling
with a ``physical'' scale choice $\mu=Q$ (${\overline{MS}}$PS). Instead we want to resum logarithms
to all-orders in the ${\rho}({\cal{R}})$ function (ECH).
The form of the ${\rho}_{n}$ RS-invariants (Eq.(4))
means that the ${\rho}_{n}$ have the structure
\begin{equation}
\rho_n = \rho_n^{\rm LL} L^n + \rho_n^{\rm NLL} L^{n-1} + \cdots\;.
\end{equation}
One can then define all-orders RS-invariant $LL$ and $NLL$ approximations to
${\rho}({\cal{R}})$, 
\begin{eqnarray}
\rho_{\rm LL}({\cal{R}})& = &-b{\cal{R}}^{2} (1 + c{\cal{R}} + \sum_{n=2}^{\infty} \rho_n^{\rm LL} L^n {\cal{R}}^{n}) 
\nonumber \\
\rho_{\rm NLL}({\cal{R}})& = &-b {\cal{R}}^{2} (1 + c {\cal{R}} + \sum_{n=2}^{\infty} (\rho_n^{\rm LL} L^n
+ \rho_n^{\rm NLL} L^{n-1}){\cal{R}}^{n} )\;.
\end{eqnarray}
The resummed ${\rho}_{\rm NLL}({\cal{R}})$ can then be used to solve for ${\cal{R}}_{\rm NLL}$ by
inserting it in Eq.(5). Notice that since 
${\Lambda}_{\cal{R}}$ involves the {\it exact} value of ${r}_{1}(1,\overline{MS})$ there
is no matching problem as in the standard $\overline{MS}$PS approach.
The resummed ${\rho}_{LL}({\cal{R}})$ can be straightforwardly numerically computed using
\begin{equation}
{\rho}_{\rm LL}(x) = \beta(a) \frac{d\cal{R}_{\rm LL}}{da} = -ba^2 \frac{d\cal{R}_{\rm LL}}{da}\;,
\end{equation}
with $a$ chosen so that ${\cal{R}}_{\rm LL}(a)=x$. The same relation
with ${\beta}(a)=-b{a}^{2}(1+ca)$ suffices for ${\rho}_{NLL}({\cal{R}})$, although in this case
one needs to remove $NNLL$ terms, e.g. an ${L}^{0}$ term which would otherwise be included in
${\rho}_{2}$. This can be accomplished by numerically taking limits ${L}\rightarrow{\infty}$
with ${L}{\cal{R}}$ fixed. \\

As we have noted a crucial feature of the effective charge approach is that it resums to all-orders
{\it RG-Predictable} pieces of the higher-order coefficients, thus the NLO ECH result (assuming $c=0$ for
simplicity) corresponds to an RS-invariant resummation (c.f. Eq.(13).) 
\begin{equation}
a+{r}_{1}{a}^{2}+{r}_{1}^{2}{a}^{3}+\cdots+{r}_{1}^{n}{a}^{n+1}+\cdots\;.
\end{equation}
Thus even at fixed-order without any resummation of large logs in ${\rho}({\cal{R}})$ a {\it partial}
resummation of large logs is automatically performed. Furthermore one might expect that the
LL ECH result contains already NLL pieces of the standard ${\overline{MS}}$PS result.\\
\begin{figure}
\begin{center}
\includegraphics[scale=0.6]{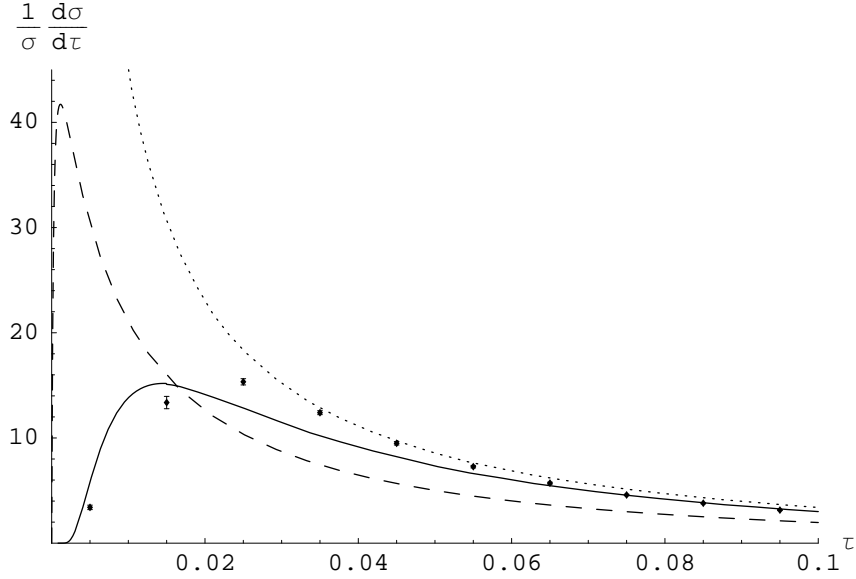}
\end{center}
\caption{Comparison of the 1-thrust distribution using various NLO approximations
in the 2-jet region. The solid curve arises from exponentiating the NLO ECH. The dashed curve
is obtained by expanding this to NLO in ${\overline{MS}}$PS. The dotted curve is an unexponentiated
NLO ECH fit. DELPHI data at $Q={M}_{Z}$ are plotted. ${\Lambda}_{\overline{MS}}=212\;\rm{MeV}$ is assumed.}
\end{figure}
\begin{figure}
\begin{center}
\includegraphics[scale=0.6]{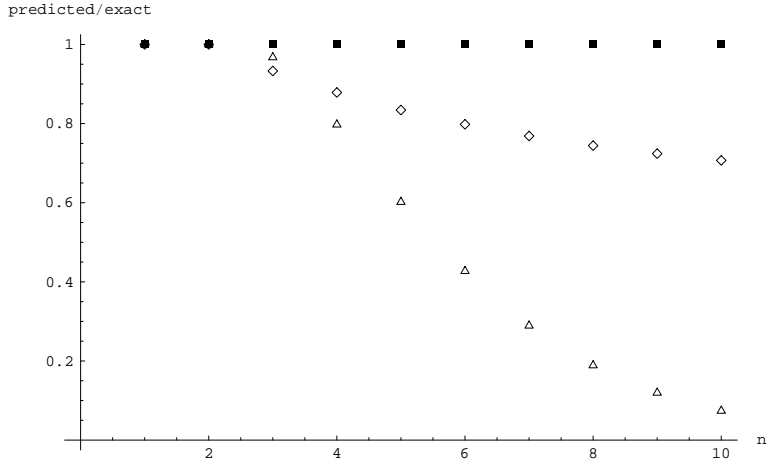}
\end{center}
\caption{For 1-thrust the ratio of the NLL ${\overline{MS}}$PS coefficient at O(${a}^{n}$) ``predicted'' from the
LL ECH result to the exact result (diamonds). The triangles show the ``prediction'' from the NLO ECH result.}
\end{figure}

In Figure 3 we show various NLO approximations. Notice that the solid curve, which
corresponds to the exponentiated NLO ECH result, is a surprisingly good fit even in
the 2-jet region, whereas the dashed curve which is the NLO $\overline{MS}$PS result, has
a badly misplaced peak. The all-orders partial resummation of large logs in Eq.(15) gives
a reasonable 2-jet peak. Figure 4 shows that the NLL ${\overline{MS}}$PS  coefficients
``predicted'' from the LL ECH result by re-expanding it in the $\overline{MS}$PS coupling
are in good agreement with the exact coeffiecients out to O(${a}^{10}$).\\


\section{Fits for ${\Lambda}_{\overline{MS}}$ and power corrections}
\noindent We now turn to fits simultaneously extracting ${\Lambda}_{\overline{MS}}$ and the size of power corrections
${C}_{1}/Q$ from the data. To facilitate this we use the result that inclusion of power
corrections effectively shifts the event shape distributions, which can be motivated
by considering simple models of hadronization, or through a renormalon analysis \cite{r8}. Thus
we define
\begin{equation}
{R}_{PC}(y)={R}_{PT}(y-{C}_{1}/Q)\;.
\end{equation}
This shifted result is then fitted to the data for 1-thrust and heavy jet mass. ${e}^{+}{e}^{-}$
data spanning the c.m. energy range from $44-189$ GeV was used (see \cite{r1} for the complete list
of references). The resulting fits for 1-thrust and heavy-jet mass are shown in Figures 5. and 6..\\

\noindent The ECH fits for thrust and heavy jet mass show great stability going from
NLO to LL to NLL, presumably because at each stage a partial resummation of higher logs is
automatically performed. The power corrections required with ECH are somewhat smaller than those found
with ${\overline{MS}}$PS, but we do not find as dramatic a reduction as DELPHI find for the means.
This may be because their analysis corrects the data for bottom quark mass effects which we
have ignored.
The fitted value of ${\Lambda}_{\overline{MS}}$ for ECH is much smaller than
that found with ${\overline{MS}}$PS, (${\alpha}_{s}(M_Z)=0.106$ (thrust) and  $0.109$ (heavy-jet
mass)). Similarly small values are found with the Dressed Gluon Exponentiation
(DGE) approach \cite{r9}.
A problem with the effective charge resummations is that the ${\rho}({\cal{R}})$
function contains a branch cut which limits how far into the 2-jet region one can go. We are
limited to $1-T>0.05{M}_{Z}/Q$ in the fits we have performed. This branch cut mirrors a corresponding
branch cut in the resummed $g_1(aL)$ function.
Similarly as $1-T$ approaches $1/3$ the leading
coefficient ${r}_{0}(L)$ vanishes and the Effective Charge formalism breaks down. We need to
restrict the fits to $1-T<0.18$. From the ``RG-predictability'' arguments we might expect
that these difficulties would also become apparent for a NNLL ${\overline{MS}}$PS resummation. One will be able to check this expectation when a result
for ${g}_{3}(a\pi L)$ becomes available.  
\begin{figure}
\begin{center}
\includegraphics[scale=0.4]{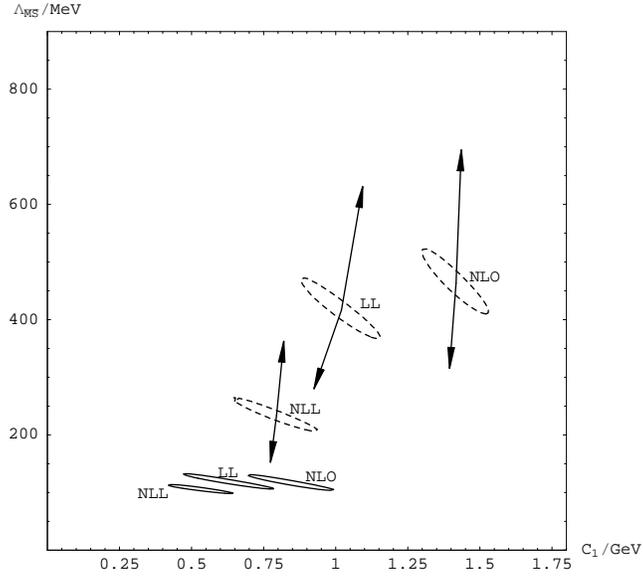}
\end{center}
\caption{Fits to 1-thrust for ${\Lambda}_{\overline{MS}}$ and $C_1$. Solid $2\sigma$ error
ellipses are for ECH, dashed are ${\overline{MS}}$PS. The arrows show the effect of varying the scale between $Q/2<\mu<2Q$.}
\end{figure}
\begin{figure}
\begin{center}
\includegraphics[scale=0.4]{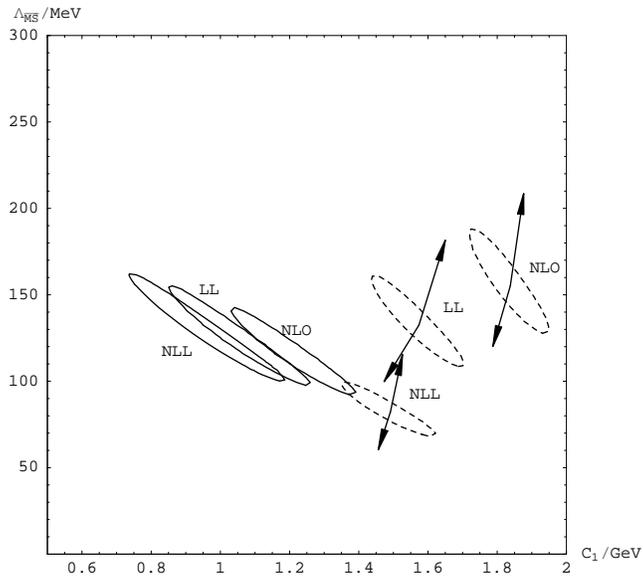}
\end{center}
\caption{Fits for heavy-jet mass.}
\end{figure}
\section{Extension to event shape means at HERA}
Event shape means have also been studied in DIS at HERA \cite{r10}.
For such processes one has a convolution of proton pdf's
and hard scattering cross-sections,
\begin{equation}
\frac{d{\sigma}(ep\rightarrow X,Q)}{dX}=\sum_{a}\int{d\xi}{f}_{a}(\xi,M)\frac{d{\hat{\sigma}}(e a\rightarrow X,Q,M)}{dX}\;. 
\end{equation}
There is no way to directly relate such quantities to effective charges. The DIS cross-sections will
depend on a {\it factorization scale} $M$, and a renormalization scale $\mu$ at NLO. In principle one could
identify unphysical scheme-dependent ${\ln}(M/{\tilde{\Lambda}}_{\overline{MS}})$ and
${\ln}({\mu}/{\tilde{\Lambda}}_{\overline{MS}})$, and physical UV $Q$-logs, and then by all-orders
resummation get the $M$ and $\mu$-dependence to ``eat itself''.
The pattern of logs is far
more complicated than the geometrical progression in the effective charge case, and a CORGI
result for DIS has not been derived so far. Instead one can use the Principle of Minimal
Sensitivity (PMS) \cite{r5}, and for an event shape mean
$\langle{y}\rangle$ look for a stationary saddle point in the $(\mu,M)$ plane \cite{r11}. It turns that
there are large cancellations between the NLO corrections for quark and gluon initiated
subprocesses. One can distinguish between two approaches, ${PMS}_{1}$ where one seeks a saddle
point in the $(\mu,M)$ plane for the sum of parton subprocesses, and ${PMS}_{2}$ where one
introduces two separate scales ${\mu}_{q}$ and ${\mu}_{g}$ and finds a saddle point
in $({\mu}_{q},{\mu}_g,M)$. ${PMS}_{1}$ gives power corrections fits comparable to
${\overline{MS}}$PS with $M={\mu}=Q$. ${PMS}_{2}$ in contrast gives substantially reduced
power corrections. This is shown in Figure 7 for a selection of HERA event shape means. 
Given large cancellations of NLO corrections RG-improvement
should be performed {\it separately} for the $q$ and $g$-initiated subprocesses, and so ${PMS}_{2}$ which indeed fits the data best, is to be preferred.  
\begin{figure}
\begin{center}
\includegraphics[scale=1.0]{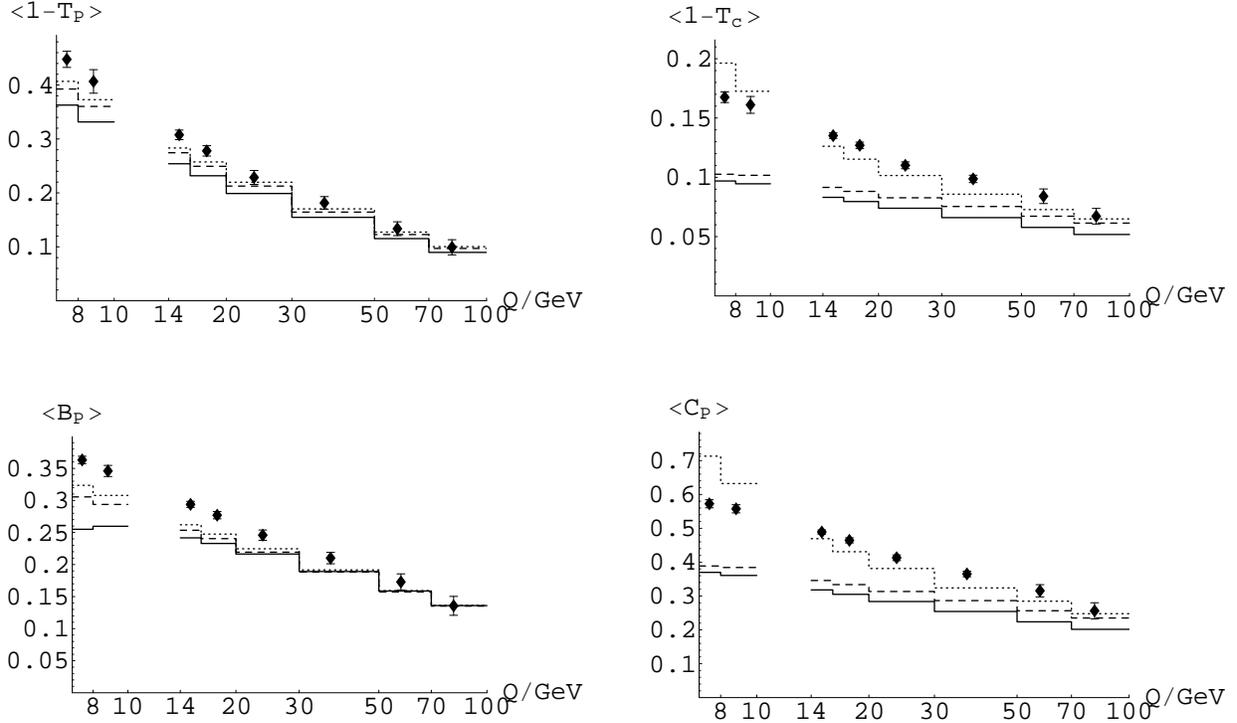}
\end{center}
\caption{The dashed line corresponds to ${PMS}_{1}$, and the solid line to
the physical scale choice $M=\mu=Q$. The dotted line is ${PMS}_{2}$ and is
in much better agreement with the data points. \cite{r11}}
\end{figure}
\section{Conclusions}
Event shape means in ${e}^{+}{e}^{-}$ annihilation are well-fitted by NLO
perturbation theory in the effective charge approach, without any power corrections being required. With the usual ${\overline{MS}}$PS approach power corrections $C_1/Q$ are required with $C_1\sim{1}$ GeV. Similarly sized power corrections are predicted in the model of Ref.\cite{r4}. It would be interesting
to modify this model so that its perturbative component matched the effective charge prediction, but this has not been done. We
showed how resummation of large logarithms in the effective charge beta-function $\rho({\cal{R}})$ could be carried out for ${e}^{+}{e}^{-}$
event shape distrtibutions. If the distributions are represented by an exponentiated effective charge then even at NLO a partial
resummation of large logarithms is performed. As shown in Figure 3 this results in good fits to the 1-thrust distribution, with
the peak in the 2-jet region in rough agreement with the data. In contrast the ${\overline{MS}}$PS prediction has a badly misplaced
peak in the 2-jet region, and is well below the data for the realistic value of ${\Lambda}_{\overline{MS}}=212$ MeV assumed. We
further showed in Figure 4 that the LL ECH result contains already a large part of the NLL ${\overline{MS}}$PS result. We found
unfortunately that $\rho({\cal{R}})$ contains a branch point mirroring that in the resummed ${g}_{1}(aL)$ function. This limited
the fit range we could consider. We fitted for power corrections and ${\Lambda}_{\overline{MS}}$ to the 1-thrust distribution
and heavy-jet mass distributions, finding somewhat reduced power corrections for the ECH fits compared to ${\overline{MS}}$PS, with
good stability going from NLO to LL to NLL. The suggestion of the ``RG-predictability'' manifested in Figure 4 would be that
the NLL ECH result contains a large part of the NNLL ${\overline{MS}}$PS result. This suggests that the branch point problem
which limits the ability to describe the 2-jet peak, would also show up given a NNLL analysis. This can be checked once the ${g}_{3}(aL)$
function becomes available. Recent work on event shape means in DIS was briefly mentioned and seemed to indicate that greatly
reduced power corrections are found when a correctly optimised PMS approach is used.

\section*{Acknowledgements}
Mrinal Dasgupta, Yuri Dokshitzer and Gavin Salam are thanked for their
painstaking organisation of this stimulating and productive workshop.

\end{document}